\documentclass[iop]{emulateapj}
\usepackage[latin1]{inputenc}
\usepackage[english]{babel}
\usepackage{amsmath}
\usepackage{amssymb}
\usepackage{latexsym}
\usepackage{epsfig}
\usepackage{subfigure}
\usepackage{natbib}
\usepackage{setspace}
\usepackage{graphicx}
\usepackage{longtable}
\usepackage{color}

\shorttitle{Polarimetry microlensing of close-in planetary
systems}\shortauthors{Sajadian \& Hundertmark}
\begin{document}

\title{Polarimetry microlensing of close-in planetary systems}
\author{Sedighe Sajadian \altaffilmark{1}, Markus Hundertmark \altaffilmark{2,3}}
\altaffiltext{1}{Department of Physics, Isfahan University of
Technology, Isfahan 84156-83111, Iran}
\altaffiltext{2}{Astronomisches Rechen-Institut, Zentrum f{\"u}r
Astronomie der Universit{\"a}t Heidelberg (ZAH), 69120 Heidelberg,
Germany} \altaffiltext{3}{Niels Bohr Institute \& Centre for Star
and Planet Formation, University of Copenhagen {\O}ster Voldgade
5,1350-Copenhagen,Denmark}\email{s.sajadian@cc.iut.ac.ir}

\begin{abstract}
A close-in giant planetary (CGP) system has a net polarization
signal whose value varies depending on the orbital phase of the
planet. This polarization signal is either caused by the stellar
occultation or by reflected starlight from the surface of the
orbiting planet. When the CGP system is located in the Galactic
bulge, its polarization signal becomes too weak to be measured
directly. One method for detecting and characterizing these weak
polarization signatures due to distant CGP systems is gravitational
microlensing. In this work, we focus on potential polarimetric
observations of highly-magnified microlensing events of CGP systems.
When the lens is passing directly in front of the source star with
its planetary companion, the polarimetric signature caused by the
transiting planet is magnified. As a result some distinct features
in the polarimetry and light curves are produced. In the same way
microlensing amplifies the reflection-induced polarization signal.
While the planet-induced perturbations are magnified, whenever these
polarimetric or photometric deviations vanish for a moment the
corresponding magnification factor or the polarization component(s)
is equal to the related one due to the planet itself. Finding these
exact times in the planet-induced perturbations helps us to
characterize the planet. In order to evaluate the observability of
such systems through polarimetric or photometric observations of
high-magnification microlensing events, we simulate these events by
considering confirmed CGP systems as their source stars and conclude
that the efficiency for detecting the planet-induced signal with the
state-of-the-art polarimetric instrument (FORS2/VLT) is less than
$\mathrm{0.1\%}$. Consequently, these planet-induced polarimetry
perturbations can likely be detected under favorable conditions by
high-resolution and short-cadence polarimeters of the next
generation.
\end{abstract}

\section{Introduction}\label{two}
The scattering of photons in stellar atmospheres creates linearly
polarized light. Due to the circular symmetry of the surface of
stars, the net polarization signal is zero
\citep{chandrasekhar60,sob}. In various types of stars, different
physical scattering mechanisms can produce a non-vanishing local
polarization signal (see e.g. Ingrosso et al. 2012). In the hot
early-type stars, for instance, Thompson scattering in their
atmospheres generates polarized light \citep{chandrasekhar60}. For
late-type main-sequence stars, the polarization is caused by both
Rayleigh scattering on neutral hydrogens and to a lesser degree by
Thompson scattering by free electrons \citep{Fluri1999,Stanflo2005}.
In cool giant stars, Rayleigh scattering on atomic and molecular
species or on dust grains generates the polarization.

During a microlensing event, the circular symmetry of images breaks
causing a time-dependent net polarization signal for the source star
\citep{schneider87,simmons95a,Bogdanov96}. The exact value of this
polarization depends on the minimum distance between lens and source
center with respect to the source radius in addition to the source
type. It reaches its maximum value when
$\mathrm{u\approx0.96\rho_{\star}}$ in microlensing events caused by
single microlenses, where $\mathrm{u}$ is the projected lens-source
distance normalized to the Einstein radius and
$\mathrm{\rho_{\star}}$ is the source radius projected on the lens
plane and normalized to the Einstein radius. The Einstein radius is
the radius of the ring of images when the lens, source and observer
are completely aligned. For early-type source stars this maximal
contribution is $\mathrm{0.6-0.7\%}$ which is measurable with
state-of-the-art polarimeters, such as the FOcal Reducer and low
dispersion Spectrograph (FORS2) polarimeter at Very Large Telescope
(VLT) telescope.

Measuring the polarization signal during microlensing events helps
to constrain the finite source size, the Einstein radius and the
limb-darkening parameters of the source star surface
\citep{yoshida06,Agol96,schneider87}. Moreover, lensing can magnify
the intrinsic polarization of source stars and make them detectable
through polarimetric microlensing. Different kinds of anomalies on
the surface or in the atmosphere of a source star can lead to a net
polarization. Magnetic fields those are distributed over the surface
of a source star can cause polarization through two channels of the
Zeeman effect and breaking the circular symmetry of the source
surface brightness due to their temperature contrasts. Photometric
observations are more efficient in detecting stellar spots than
polarimetric observations, but by using polarimetric observations we
can estimate the magnetic field of source star spots
\citep{sajadian2015a}. Circumstellar hot disks around the
microlensed stars located at the Galactic bulge lead to net
polarization signals for their host stars due to their projected
elliptical shapes. Indeed, these distant disks are unresolvable from
their host stars. Polarimetric microlensing observations of these
systems can offer a lower limit to the disk inner radius
\citep{sajadian2015b}. In addition, rapid stellar rotation makes (i)
stars oblate which breaks the spherical symmetry of the source and
(ii) the gravity-darkening effect which breaks the respective
circular symmetry of the stellar surface brightness. Hence, stellar
rotation can also lead to a net polarization depending on the
inclination angle of its rotational axis as well as its angular
speed. The corresponding stellar ellipticity causes some time shifts
in the peak positions of the polarimetry and light curves of
high-magnification microlensing events. Measuring these shifts can
only viably achieved by using polarimetric microlensing
\citep{sajadian2016}.

A distant CGP system has also a net polarization signal which varies
with time while the planet is orbiting its host star. This
polarization signal can be caused by (i) the reflection of the
stellar light from the planet surface (e.g. Seager et al. 2000),
(ii) the stellar occultation by the transiting planet (e.g. Carciofi
\& Magalh\~{a}es 2005) or (iii) the planet occultation by the source
edges while the planet is occulted by its host star. The
polarization signal due to the reflection is maximal when the phase
angle is about $\mathrm{90}$ degree, where the phase angle is
measured between the source and Earth as seen from the planet center
and the occultation polarization is maximal when the planet is
crossing the source edges. As a side-remark, the polarization caused
by the occultation of the planet by the star is maximal in the
infrared bands and its duration is too short. Generally, the total
polarization signal due to a distant CGP system is too low to be
directly measured. Again, microlensing can magnify all these effects
of close-in giant planets orbiting source stars beyond the detection
limit.

The detectability of such CGP systems in microlensing events through
pure photometric observations were studied in great details. For
instance, Graff \& Gaudi (2000) proposed detecting Jupiter-size
planets around the source stars during caustic-crossing microlensing
events. This aspect was further addressed by Sajadian \& Rahvar
(2010) who studied the efficiency of detecting such systems as a
function of wavelength. Microlensing light curves of transiting
planets orbiting source stars were studied by Rybicki \& Wyrzykowski
(2014) and they have concluded that the probability for detecting a
transiting planet in a given microlensing event is of the order of
$\mathrm{2\times 10^{-6}}$. We propose \emph{polarimetric
observations} as a new channel for detecting and even characterizing
such systems in the Galactic bulge through high-magnification or
caustic-crossing microlensing events. The outline of this paper is
as follows: in the next section, we first outline our framework for
describing a planetary orbit and its projection on the sky for
applying the lensing formalism. Then, we explain how to calculate
the net polarization signal of a CGP system. In section
(\ref{four}), we study the polarimetric microlensing signatures of
these systems. Finally, we discuss on probability of detecting CGP
systems through photometric or polarimetric observations of
high-magnification caustic-crossing microlensing events in section
(\ref{five}).

\section{Polarization signal of a distant CGP system}\label{three}
In this section, we outline the formalism for calculating the
polarization signal of a CGP system. First, we introduce the
relevant parameters and coordinate systems to establish a planetary
orbit. Then, we assess different contributions to the net
polarization signal of a CGP system.

\subsection{Kepler parameters for a planetary orbit}
The Kepler problem of a planetary system whose components interact
through their gravitational force is converted to motion of an
object with the reduced mass
$\mathrm{\mu=M_{\star}M_{p}/(M_{\star}+M_{p})}$ around their center
of mass over an ellipse, where $\mathrm{M_{p}}$ and
$\mathrm{M_{\star}}$ are the masses of the planet and its parent
star respectively. The radial distance of the reduced mass relative
to the center of mass is given by (e.g. Landau \& Lifshitz 1969,
Smart 1980):
\begin{eqnarray}
r(\xi)=a(1-e \cos\xi),
\end{eqnarray}
where $\mathrm{a}$ is the semi-major axis and $\mathrm{e}$ is the
orbital eccentricity. $\mathrm{\xi}$ is called the true anomaly and
its dependence on time is given by the Kepler equation:
\begin{eqnarray}
\xi-e \sin\xi= \omega(t-t_{p})=\phi,
\end{eqnarray}
where $\mathrm{\omega=2\pi/T}$ is the orbital angular velocity,
$\mathrm{T=2\pi\sqrt{a^3/G(M_{\star}+M_{p})}}$ is the orbital period
and given by the Kepler's third law and $\mathrm{t_{p}}$ is the time
of the perihelion passage. The Kepler equation can be numerically
solved by using a series expansion of the Bessel functions
\citep{Watson1966,Nucita2014}:
\begin{eqnarray}
\xi=\phi+ \sum_{n=1}^{+\infty}\frac{2}{n}J_{n}(ne)\sin(n\phi),
\end{eqnarray}
where $\mathrm{J_{n}(ne)}$ is the Bessel function of the order
$\mathrm{n}$. The radial motion of the planet around its host star
is given by $\mathrm{r_{p}=r(\xi)\mu/M_{p}}$.

We define a Cartesian coordinate system
$\mathrm{(x_{p},y_{p},z_{p})}$ with the origin at the center of the
source star. Its $\mathrm{y_{p}}$ and $\mathrm{z_{p}}$-axes coincide
with the orbital plane and point towards the semi-major and minor
axes respectively. The $\mathrm{x_{p}}$-axis is orthogonal to this
plane, so that all basis vectors form a right-handed system. In this
coordinate system, the components of the position of the planet
center at each time is given by (e.g. Dominik 1998):
\begin{eqnarray}
y_{c,p}&=&\frac{\mu}{M_{p}}a(cos\xi-e),\nonumber\\
z_{c,p}&=&\frac{\mu}{M_{p}}a\sqrt{1-e^2}\sin\xi,\nonumber\\
x_{c,p}&=&0.
\end{eqnarray}
We also consider an observer coordinate system describing the sky
plane $\mathrm{(x_{o},y_{o},z_{o})}$ with the origin at the center
of the source star. In this coordinate system, the $\mathrm{y_{o}}$
and $\mathrm{z_{o}}$-axes are on the sky plane and its
$\mathrm{x_{o}}$-axis points towards the observer. In order to
convert the first coordinate system to the second one, we first need
to rotate the planetary orbital plane around the
$\mathrm{x_{p}}$-axis by the angle $\mathrm{\beta}$ which is the
angle between the semi-major axis of the planetary orbit with
respect to the intersection line of the sky plane and the orbital
plane. Then, we project the planetary orbit on the sky plane by
rotating it around the semi-major axis of the planetary orbit by the
inclination angle $\mathrm{i}$ which is the angle between its
semi-minor axis and sky plane.

Based on these coordinate systems, the phase angle i.e. the angle
between the source star and the observer as seen from the planet
center, is given by:
$\mathrm{\cos\alpha=-x_{c,o}/\sqrt{x_{c,o}^{2}+y_{c,o}^{2}+z_{c,o}^{2}}}$,
where $\mathrm{(x_{c,o},y_{c,o},z_{c,o})}$ describes the position of
the planet center at each time in the observer coordinate system:
\begin{eqnarray}
y_{c,o}&=&\cos\beta~y_{c,p} + \sin\beta~z_{c,p},\nonumber\\
z_{c,o}&=&-\cos i\sin\beta~y_{c,p}+\cos i \cos\beta~z_{c,p},\nonumber\\
x_{c,o}&=&\sin i\sin\beta~y_{c,p}-\sin i \cos\beta~z_{c,p}.
\end{eqnarray}

In order to describe each point on the planet surface, we require a
third coordinate system $\mathrm{(x',y',z')}$ with the origin at the
planetary center. The $\mathrm{x'}$-axis is pointing towards the
host star, the $\mathrm{y'}$-axis is contained in the plane
$\mathrm{x_{o}-x'}$ and its $\mathrm{z'}$-axis is normal to the
plane $\mathrm{x'-y'}$, so that one gets a right-handed system. Each
point on the planet surface in this system is given by:
\begin{eqnarray}
x'&=&R_{p}\sin\theta \cos\phi, \nonumber\\
y'&=&R_{p}\sin\theta \sin\phi, \nonumber\\
z'&=&R_{p}\cos\theta,
\end{eqnarray}
where $\mathrm{R_{p}}$ is the planet radius, $\mathrm{\theta}$ and
$\mathrm{\phi}$ are the polar and azimuthal angles on the planetary
surface. In order to convert this coordinate system to the
observer's system, we first rotate the coordinate system around the
$\mathrm{z'}$-axis by the phase angle $\mathrm{\alpha}$. Then, we
rotate the resulting system around its $x'$-axis by
$\mathrm{-\gamma}$, where $\mathrm{\gamma=\cos^{-1}(
y_{c,o}/\sqrt{x_{c,o}^2+y_{c,o}^2+z_{c,o}^2})}$ is the angle between
the line pointing towards the center of the planet and the
$\mathrm{y_{o}}$-axis as seen from the host star.

\begin{figure*}
\centering
\includegraphics[angle=270,width=8.cm,clip=]{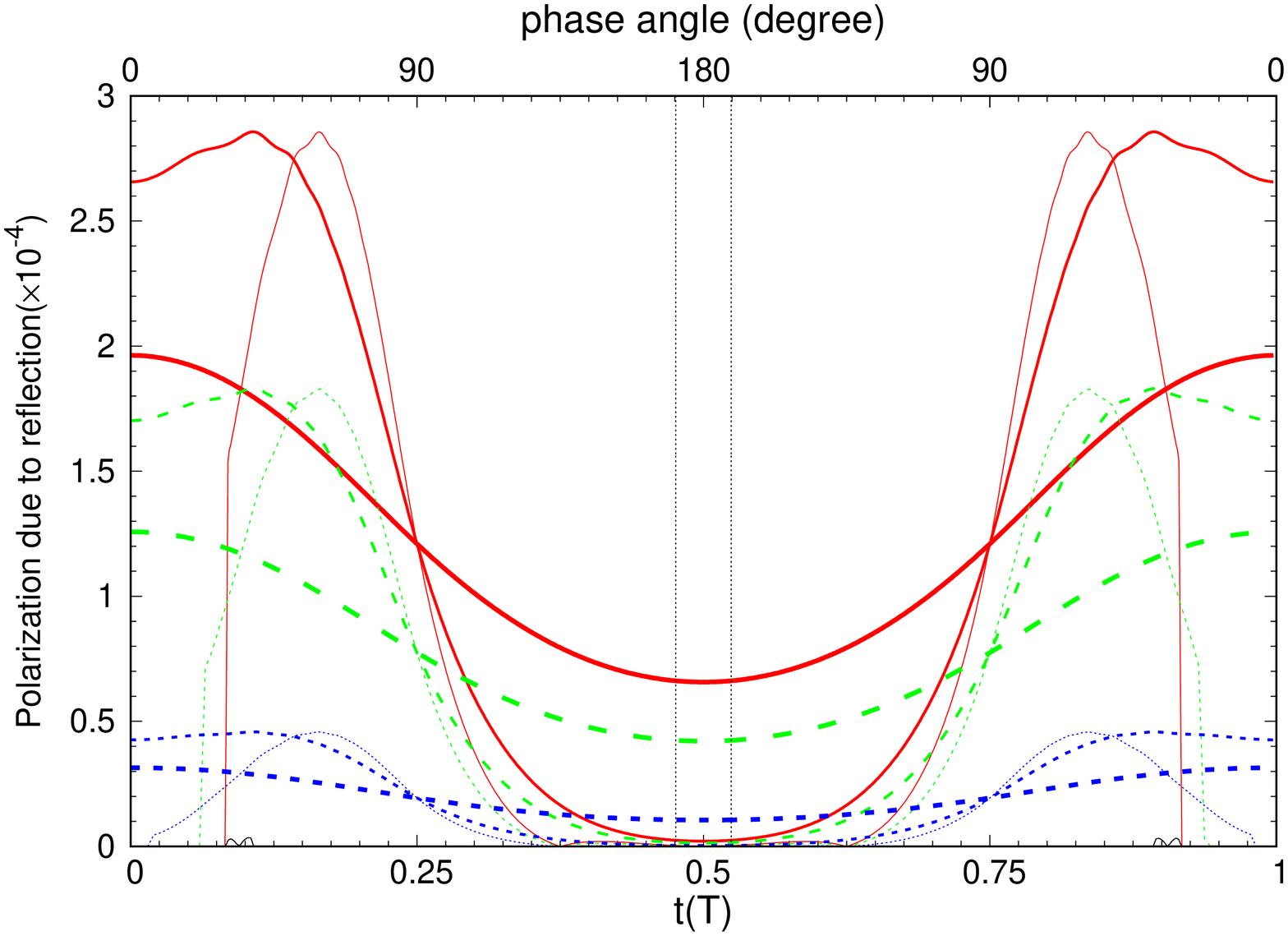}
\includegraphics[angle=270,width=8.cm,clip=]{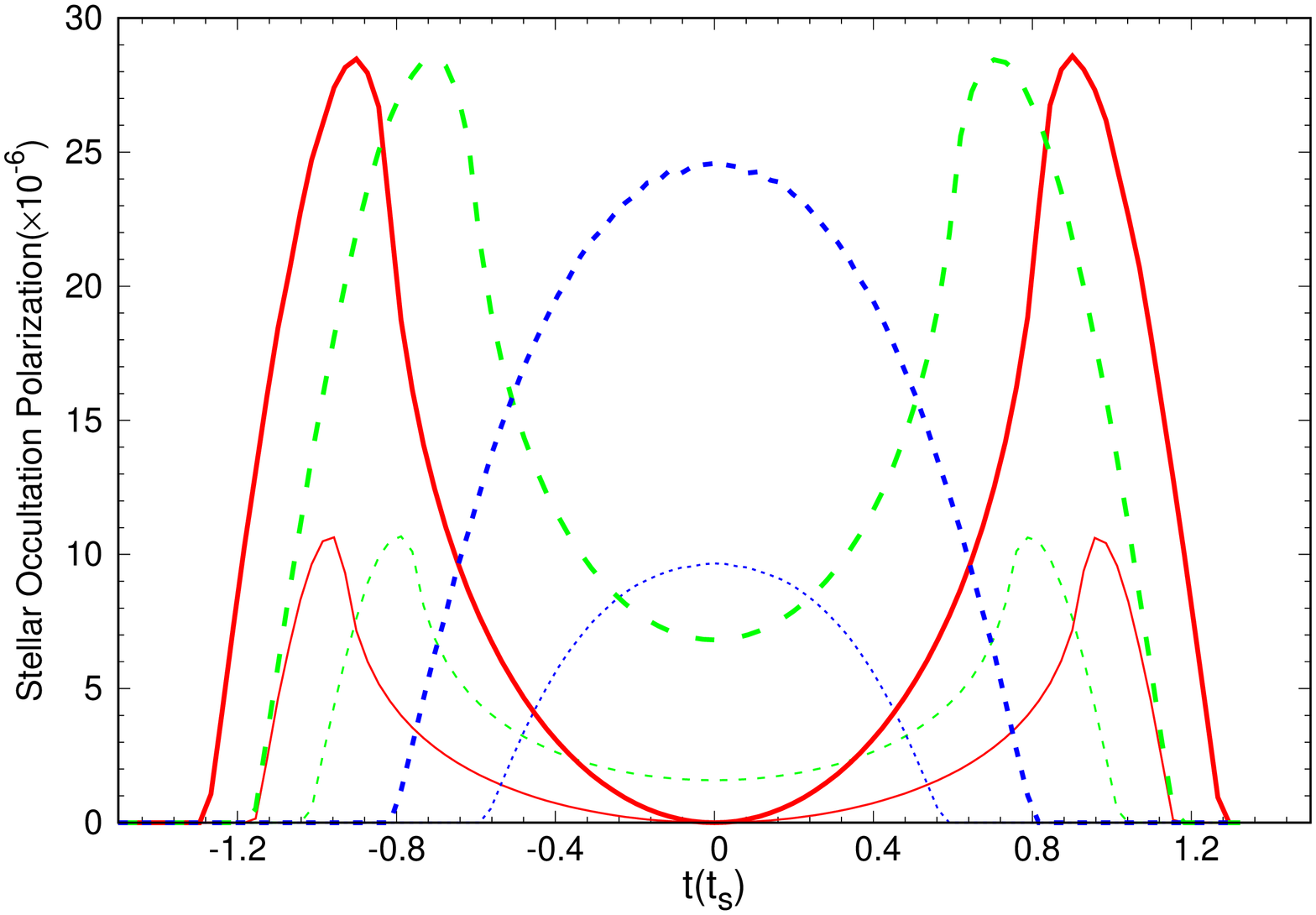}
\caption{The polarization signal of a CGP system is shown. Left
panel: the time-dependent polarization signal due to the reflected
light from the planet (normalized to the planetary orbital period
$\mathrm{T}$). The polarizations are plotted for three different
values of the planet semi-major axis $\mathrm{a=0.008~A.U.}$ (red
solid), $\mathrm{a=0.01~A.U.}$ (green dashed) and
$\mathrm{a=0.02~A.U.}$ (blue dotted) and three different inclination
angles $\mathrm{i=80,~40,~10}$ degree from thickest lines to
thinnest lines respectively. Right panel: the polarization signal
due to the occultation of the source surface by the planet versus
time (normalized to source crossing time by planet
$\mathrm{t_{s}}$). The polarimetry curves are plotted for two values
of the planet radius $\mathrm{R_{p}=R_{J}}$ (thin curves) and
$\mathrm{R_{p}=2~R_{J}}$ (thick curves) and three different
inclination angles $\mathrm{i=90}$ degree (red solid),
$\mathrm{75^{\circ}}$ (green dashed) and $63^{\circ}$ (blue
dotted).} \label{fig1}
\end{figure*}
Light coming from a CGP system located in the Galactic bulge is
linearly polarized which is due to (i) the reflection of the host
star light in the planetary atmosphere, (ii) the stellar occultation
by the transiting planet or (iii) the planetary occultation by the
source star. The latter is negligible given the faintness of the
planet itself. We will expand on the two relevant components of the
polarization in the following.

\subsection{Polarization due to reflection}
The polarization caused by reflected starlight in a CGP system is a
well-studied phenomenon. Chandrasekhar (1950,1960) and Horak \&
Chandrasekhar (1961) have analytically solved the radiative transfer
equations to calculate the intensity and the state of polarization
of the light reflected by a semi-infinite homogeneous plane-parallel
atmosphere in the context of conservative and non-conservative
Rayleigh scattering. Subsequently, many researchers have applied
Chandrasekhar's approaches to the polarimetry of exo-planets (e.g.
Natraj et al. 2009, Kane \& Gelino 2010).

Under the assumption of radiative equilibrium, for a given chemistry
and sources of absorption and scattering, Seager et al. (2000) have
presented light and polarimetry curves of a CGP system. Moreover,
many numerical models have been developed to calculate the
polarization and intensity of light reflected by planets (see e.g.
Sengupta \& Maiti 2006, Stam 2008, Buenzli \& Schmid 2009, Cahoy et
al. 2011, Kostogryz et al. 2011, Whitney 2013). Indeed, the
reflected light by a planet is significantly polarized, whereas the
direct light from the host star is generally unpolarized due to the
symmetry of the stellar surface brightness. Accordingly, polarimetry
was recommended as a method for detecting and characterizing
Jupiter-like planets \citep{Hough2003,Stam2004}.

In this work, we use the analytical framework developed by
Madhusudhan \& Burrows (2012) to calculate the Stokes parameters
\citep{Stokes1852,Tinbergen96}, which was based on the Chandrasekhar
(1950,1960)'s approach assuming a cloud-free homogeneous (in the
scattering albedo) and semi-infinite atmosphere.

The left panel of Figure (\ref{fig1}) shows the time-dependent
polarization signal due to the reflected light from the planet
surface. This figure is similar to the bottom panel of Figure (4) of
Madhusudhan \& Burrows (2012). However, the scales are different
because we normalize the polarized flux to the total flux due to the
planet and its host star, whereas in the mentioned reference the
polarized flux is normalized to the flux of the planet itself.
According to this figure, the polarization curve due to the
reflection has two peaks when $\mathrm{\alpha\sim90^{\circ}}$. The
polarization curves are plotted for three different values of the
planet semi-major axis $\mathrm{a=0.008~A.U.}$ (red solid),
$\mathrm{a=0.01~A.U.}$ (green dashed) and $\mathrm{a=0.02~A.U.}$
(red dotted) and three different inclination angles
$\mathrm{i=80,~40,~10}$ degree from thickest lines to thinnest lines
respectively. The vertical black dashed lines show the source
radius. Consequently, the polarization signal is detectable while
the planet is out of the source surface and it does not transit its
host star, except for the small inclination angles. Closer planets
lead to higher reflection polarizations. The small black features in
this figure represent the polarization signal due to the planet
occultation by its host star for $\mathrm{a=0.008~A.U.}$ and
$\mathrm{i=80^{\circ}}$. This polarization occurs whenever the
planet is behind its host star surface and in that case the
contribution is small.

\subsection{Polarization due to the stellar occultation by planet}
The occultation of the source surface by its planet breaks the
circular symmetry of the stellar surface brightness and produces a
linear polarization signal. The value of this polarization is
maximal when the planet is crossing the edge of the source
\citep{Carciofi2005,Wiktorowicz2014,Kostogryz2015}. However, the
occultation happens for phase angles greater than
$\mathrm{90^{\circ}}$ (i.e. when the planet is passing in front of
the source star) as well as when $\mathrm{R_{c,o}<R_{\star}+R_{p}}$,
where $\mathrm{R_{\star}}$ is the source radius and
$\mathrm{R_{c,o}=\sqrt{z_{c,o}^{2}+y_{c,o}^{2}}}$ is the projected
radial distance of the planet center with respect the source center.
The stellar occultation only happens when the inclination angle is
greater than: $\mathrm{i\geq \cos^{-1}(R_{s}+R_{p})/b_{p}}$, where
$\mathrm{b_{p}}$ is the semi-minor axis of the planetary orbit after
rotating the orbital plane by $\mathrm{\beta}$. Our formulism for
calculation of the occultation polarization is similar to the one
introduced by Carciofi \& Magalh\~{a}es (2005). The only difference
is that we consider a non-circular planetary orbit and as a result
we need two rotation angles to project the planetary orbit onto the
sky plane. Whereas in the aforementioned reference only circular
planetary orbits were considered which need just a single projection
angle.

A stellar occultation inevitably leads to a linear polarization. The
right panel of Figure (\ref{fig1}) represents the time-dependent
polarization signal due to the occultation of the source surface by
the planet expressed in the unit of the source star crossing time
(i.e. $\mathrm{t_{s}=T~R_{\star}/(2\pi a)}$). The polarimetry curves
are plotted for two values of the planet radius
$\mathrm{R_{p}=R_{J}}$ (thin curves) and $\mathrm{R_{p}=2 R_{J}}$
(thick curves) and three different inclination angles
$\mathrm{i=90}$ degree (red solid), $\mathrm{75^{\circ}}$ (green
dashed) and $63^{\circ}$ (blue dotted). According to this figure,
the polarimetry curve due to the occultation has one or two peaks,
which all happen when the planet is crossing the source edge. The
time interval between these peaks depends on the inclination angle.
This figure is similar to the figure (4) of Carciofi \&
Magalh\~{a}es (2005). However, the semi-major axis of the planet
orbit in our calculation is $\mathrm{0.25}$ times smaller. In
addition, we have considered different wavelengths.

\section{High-magnification microlensing of a CGP system}\label{four}
In this section, we study polarimetric microlensing events of the
CGP system that was discussed in the previous section. In these kind
of events we encounter two time-dependent properties: (i) the lens
distance from the planetary system and (ii) the position of the
planet with respect to its host star. We focuss on close-in
planetary systems, because for these systems the effect of the
reflected light from the planet surface is considerably higher and
the probability that the planet transiting the source star is high.
These CGP systems have time-dependent intrinsic polarization signals
which are likely not detectable when these systems are located in
the Galactic bulge. Lensing can magnify these low polarization
signals if (i) the the angular separation between lens and CGP
system is very small (i.e. fractions of a $\mathrm{mas}$) or (ii) if
this system passes the caustic of a multiple-lens system. In the
following subsections, we analyze how lensing changes the
occultation and reflection polarizations respectively and focus on
the first scenario.

\begin{figure*}
\centering \subfigure[] {
\includegraphics[angle=270,width=8.cm,clip=]{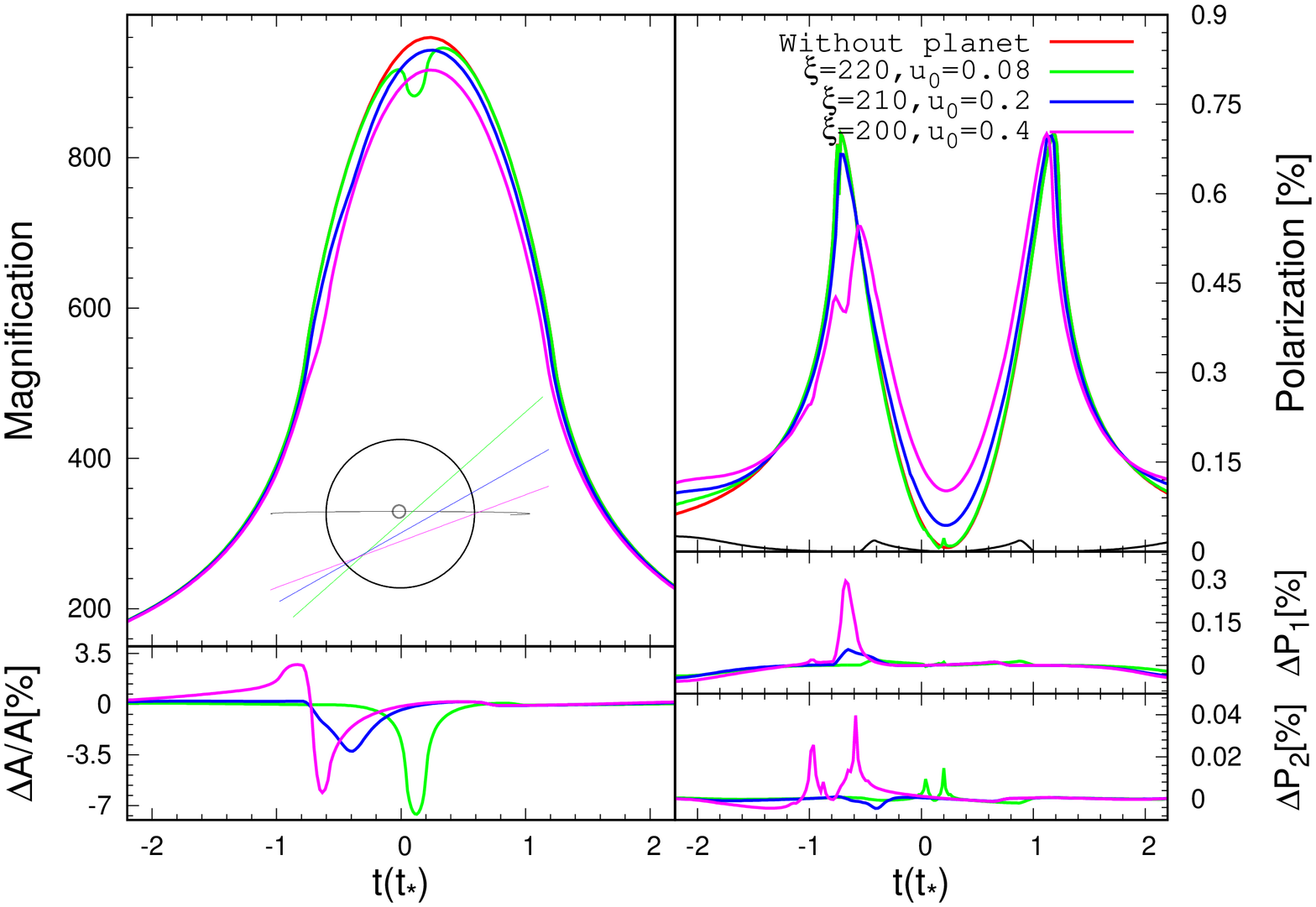}
\label{fig3a} } \subfigure[] {
\includegraphics[angle=270,width=8.cm,clip=]{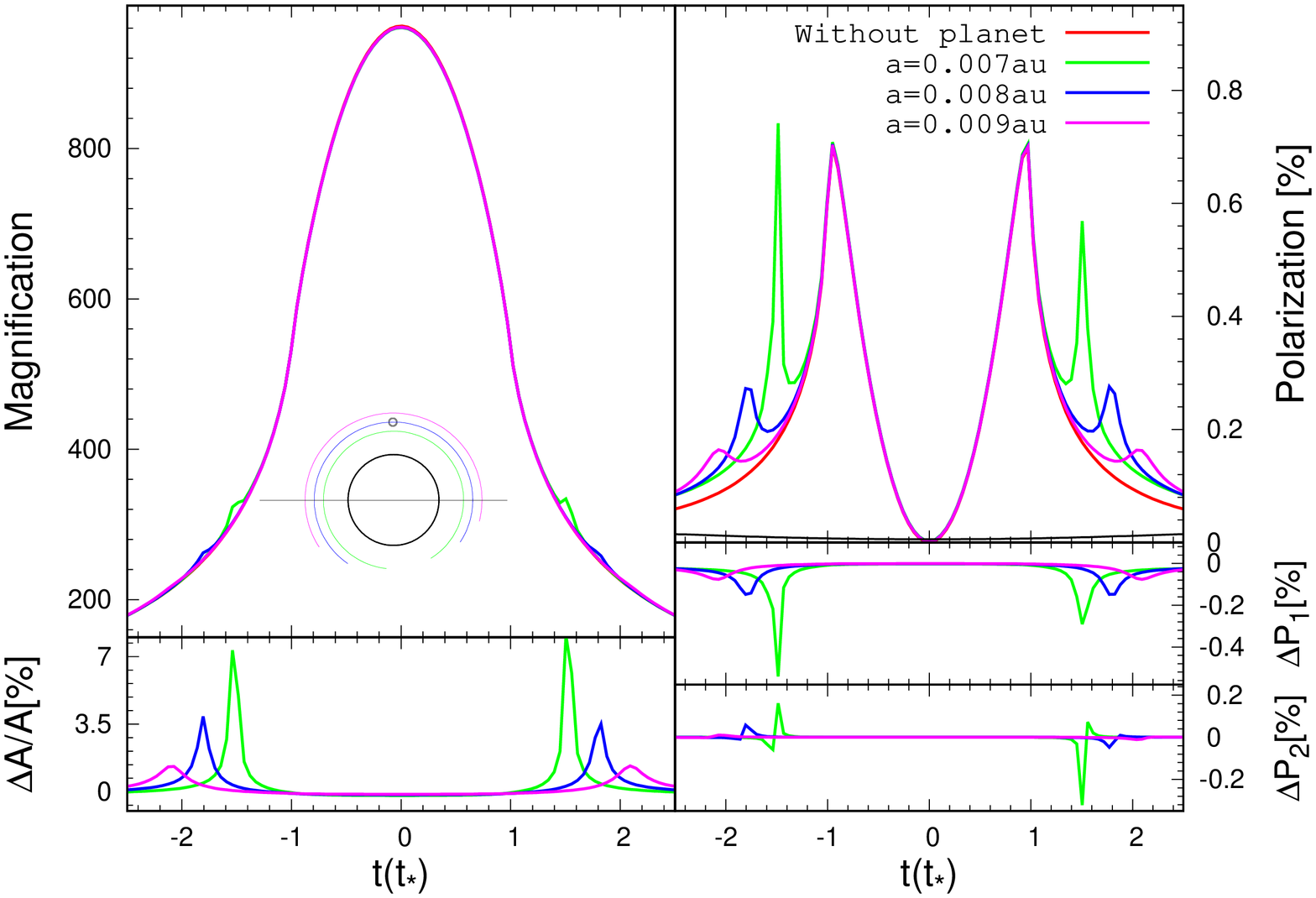}
\label{fig3b}} \subfigure[] {
\includegraphics[angle=270,width=8.cm,clip=]{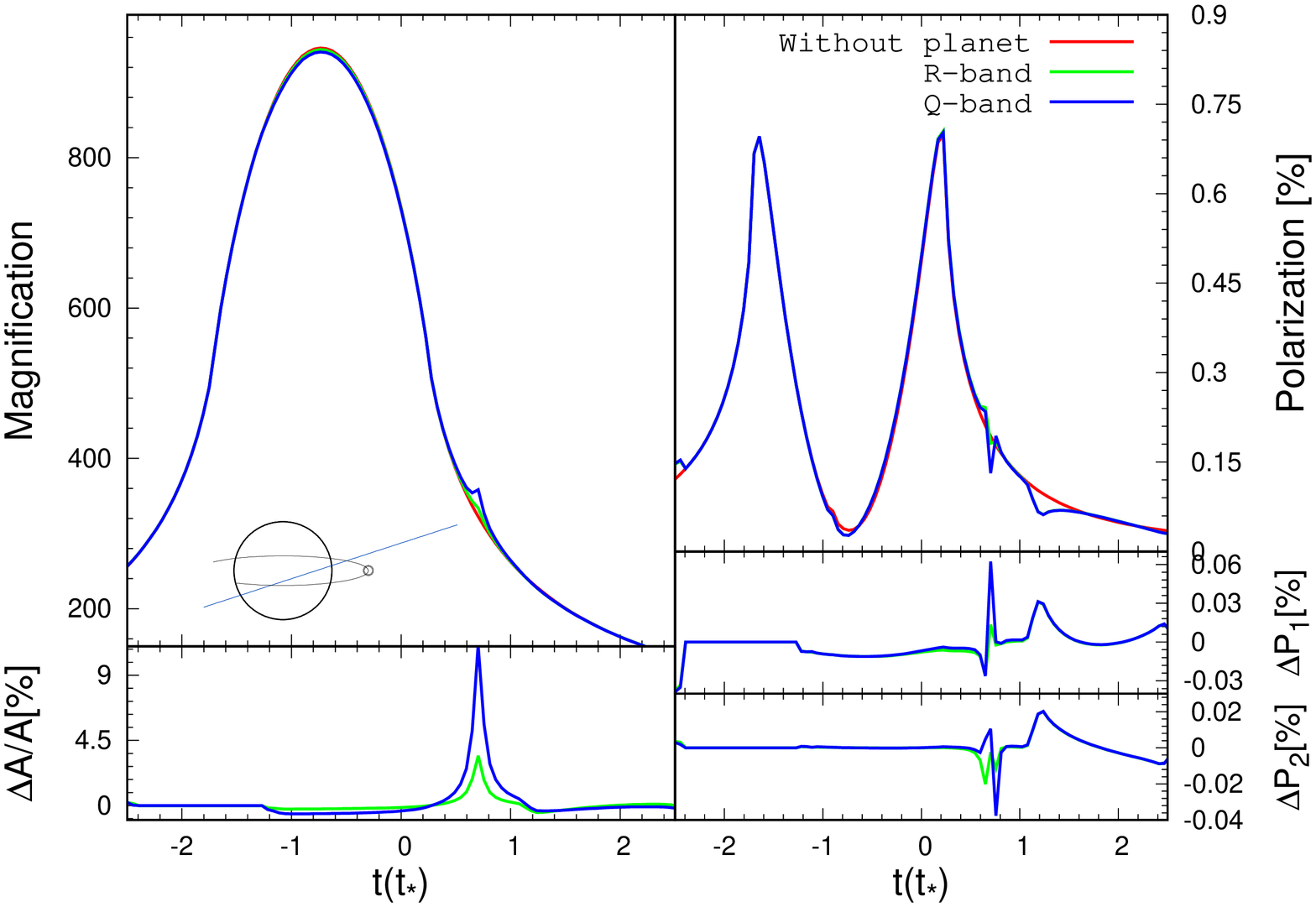}
\label{fig3c} } \subfigure[] {
\includegraphics[angle=270,width=8.cm,clip=]{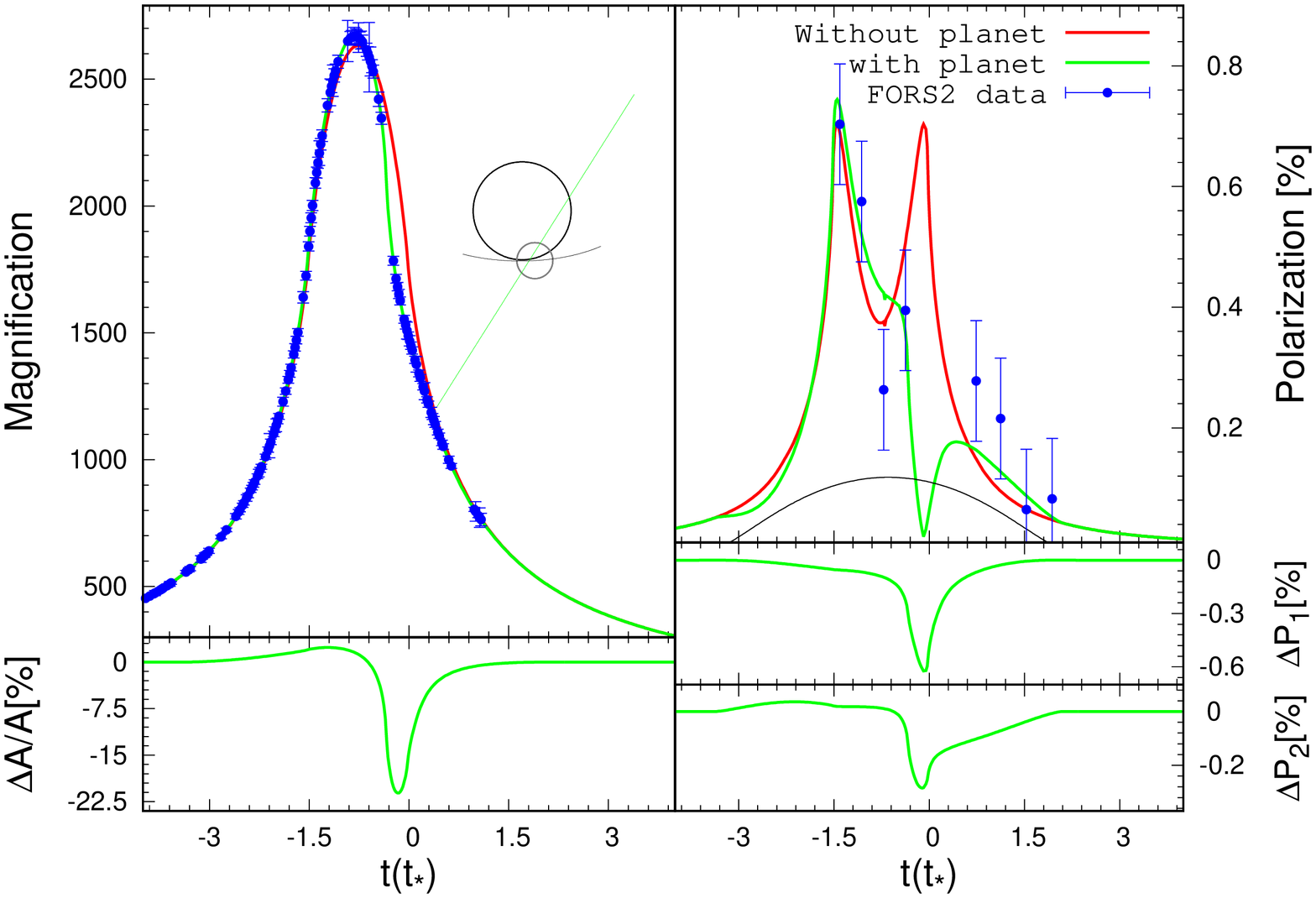}
\label{fig3d}}\caption{Typical high-magnification microlensing
events of CGP systems. In each subfigure, light and polarimetric
curves are shown in the left and right panels respectively. The
source and its planet (black and gray circles), the projected path
of the planet around its host star (solid gray lines) and the lens
trajectories (color lines) are shown in the inset in the left-hand
panels. Simple models without a planetary perturbation are plotted
with red solid lines. The black thinner curves in the right panels
represent the intrinsic polarization signals of the CGP systems
varying with time. The photometric and polarimetric residuals with
respect to the simple models are shown in the lower panels. The
parameters used to make these figures are reported in Table
(\ref{tab1}).}\label{fig2}
\end{figure*}

\subsection{Lensing of the occultation polarization}
In a high-magnification microlensing event of a CGP system in which
the lens passes the source surface, the occultation signature is
most likely magnified and causes some photometry and polarimetry
deviations in the related curves.

In order to formulate the polarization signal, we use the Stokes
parameters $\mathrm{S_{I}}$, $\mathrm{S_{Q}}$, $\mathrm{S_{U}}$ and
$\mathrm{S_{V}}$ which describe the total intensity, two components
of linear polarized intensities and the circular polarized intensity
over the source surface respectively \citep{Tinbergen96}. The
scatting process produces linear polarizations on the source
surface, so we set $\mathrm{S_{V}=0}$. The other Stokes parameters
are calculated by integrating of the total and polarized flux over
the source surface and by considering the magnification factor for
each element of the source surface:
\begin{eqnarray}
S_{I,\star}&=&~\int_{0}^{1}\rho~d\rho\int_{-\pi}^{\pi}d\phi F^{\star}(\mu)~ A(u),\\
\left( \begin{array}{c}
S_{Q,\star}\\
S_{U,\star}\end{array}\right)&=&\int_{0}^1\rho~d\rho\int_{-\pi}^{\pi}d\phi F^{\star}_{p}(\mu) A(u) \left( \begin{array}{c} -\cos 2\phi \nonumber \\
\sin 2\phi \end{array} \right),
\end{eqnarray}
where $\mathrm{\rho}$ is the distance from the center to each
projected element over the source surface normalized to the source
radius, $\mathrm{\mu=\sqrt{1- \rho^{2}}}$, $\mathrm{\phi}$ is the
azimuthal angle between the lens-source connection line and the line
from the center to each element over the source surface,
$\mathrm{u}$ is the distance of each projected element over the
source surface with respect to the lens position and $A(u)$ is the
magnification factor. Noting that the Stokes parameters are
normalized to $\mathrm{R^{2}_{\star}}$. The total and polarized
fluxes $\mathrm{(F^{\star},F^{\star}_{p})}$ depend on the type of
the source star and its temperature (see e.g. Ingrosso et al.
2012,2015).

Let us assume the source star is a transiting CGP system. In that
case the total stokes parameters are divided into two parts, the
first part ($\mathrm{S_{I,\star},S_{Q,\star},S_{U,\star}}$) is the
total Stokes parameters contribution due to the source star itself
and the second part ($\mathrm{S_{I,p},S_{Q,p},S_{U,p}}$) is the
contribution by the planet which is evaluated by integrating over
the planet surface considering the total and polarized fluxes of its
host star at the planet location. Consequently, the total Stokes
parameters $\mathrm{(S'_{I},S'_{Q},S'_{U})}$ are given by:
\begin{eqnarray}
S'_{i}=R^{2}_{\star}S_{i,\star}-R^{2}_{p}S_{i,p},
\end{eqnarray}
where $i\in(I, Q, U)$. The planet Stokes parameters
$\mathrm{(S_{I,p},S_{Q,p},S_{U,p})}$ depend on the location of the
planet on the source surface with respect to the lens position and
the planet size. We assume that the observation is carried out in
the optical band, so that the thermal radiation of the planet is
negligible.

The magnification factor of such a CGP system is given by:
\begin{eqnarray}
A'=\frac{S_{I,\star}-\delta S_{I,p}}{S_{I,\star,0}-\delta
S_{I,p,0}}=A\frac{1-\delta S_{I,p}/S_{I,\star}}{1-\delta
S_{I,p,0}/S_{I,\star,0}},
\end{eqnarray}
where the index $0$ of the Stokes parameters refers to the
contribution without lensing effect and
$\mathrm{\delta=R_{p}^{2}/R_{\star}^{2}}$ represents the ratio of
the planet area to its host star area. The parameters with denoted
with a prime contain the polarimetric signal caused by the
transiting planet. As expected, the relative size of the planet to
the source star is an important factor in the magnitude of the
perturbation due to a transiting planet.

When the lens is approaching the source star surface and resides far
from the transiting planet, the planet-induced deviation in the
photometry curve is positive, because the term
$\mathrm{S_{I,\star}}$ is magnified significantly. The second term
$\mathrm{S_{I,p}}$ will be magnified whenever the lens approaches
the planet position or is crossing the planet surface resulting a
negative deviation in the light curve. Whenever the total deviation
vanishes, i.e. $\mathrm{S_{I,p}/S_{I,\star}=
S_{I,p,0}/S_{I,\star,0}}$, the magnification factor is equal to the
magnification factor of the source star without the planet and also
equals to the magnification of the planet itself
$\mathrm{A'=A=A_{p}=S_{I,p}/S_{I,p,0}}$. $\mathrm{A_{p}}$ depends on
the location of the planet with respect to the source center and its
size. Let us assume the lens position can be deduced from
photometric measurements. The magnification factor in this case
gives us some information on the planet characteristics. Generally,
depending on the lens and planet trajectories over the source
surface the planet-induced deviations can have different shapes. For
instance, they could be always negative (or always positive).

We can write the polarization vector as $\mathrm{P'=(S'_{Q}/S'_{I}
,S'_{U}/S'_{I})}$ whose components are given by:
\begin{eqnarray}
P'_{1}&=&P_{1}\frac{1-\delta~S_{Q,p}/S_{Q,\star}}{1-\delta~S_{I,p}/S_{I,\star}},\nonumber\\
P'_{2}&=&P_{2}\frac{1-\delta~S_{U,p}/S_{U,\star}}{1-\delta~S_{I,p}/S_{I,\star}}.
\end{eqnarray}
The planet-induced polarimetric perturbations can also be in the
form of positive and negative deviations. Similarly, whenever these
deviations in each polarization component vanishes while the lens is
magnifying the transiting planet effect, we will have
$\mathrm{P'_{1}=P_{1}=P_{1,p}=S_{Q,p}/S_{I,p}}$ or
$\mathrm{P'_{2}=P_{2}=P_{2,p}=S_{U,p}/S_{I,p}}$. The value of that
polarization component at that time gives us one constraint on the
 properties of the planet.
\begin{deluxetable*}{ccccccccc}
\tablecolumns{9} \centering \tablewidth{0.8\textwidth}
\tabletypesize \footnotesize \tablecaption{The parameters of
microlensing events shown in Figures \ref{fig3a}, \ref{fig3b},
\ref{fig3c} and \ref{fig3d}.} \tablehead{ \colhead{} &
\colhead{$\mathrm{M_{p}(M_{J})}$}&
\colhead{$\mathrm{R_{p}(R_{J})}$}& \colhead{$\mathrm{a(A.U.)}$} &
\colhead{$\mathrm{i^{\circ}}$}&
\colhead{$\mathrm{u_{0}(\rho_{\star})}$}&
\colhead{$\mathrm{\xi^{\circ}}$}&
\colhead{$\mathrm{t_{0}(t_{\star})}$}&
\colhead{$\mathrm{t_{p}(T)}$}}& \startdata
$2(a)$ & $0.4$ & $0.9$ & $0.008$  & $89$ & $--$ & $--$ & $0.23$& $-0.25$\\
\\ 
$2(b)$ & $0.4$ & $0.9$ & $--$ & $10$ & $0.009$ & $180$ & $0$ & $-0.25$\\
\\ 
$2(c)$ & $0.5$ & $1.0$ & $0.008$  & $80$ & $-0.2$ & $18$ & $-0.73$ & $0$\\
\\ 
$2(d)$ & $0.68$ & $1.2$ & $0.05$  & $41$ & $0.6$ & $296$ & $-0.46$ & $-0.24$\\
\enddata
\tablecomments{The columns contains (i) the figure, (ii) the planet
mass $\mathrm{M_{p}(M_{J})}$, (iii) the planet radius
$\mathrm{R_{p}(R_{J})}$, (iv) the semi-major axis
$\mathrm{a(A.U.)}$, (v) the inclination angle of the planetary
orbital plane with respect to the sky $\mathrm{i^{\circ}}$, (vi) the
impact parameter of the lens trajectory with respect to the source
center normalized to $\mathrm{\rho_{\star}}$
$\mathrm{u_{0}(\rho_{\star})}$, (vii) the angle of the lens
trajectory with respect to the semi-major axis of the planetary
orbit $\mathrm{\xi^{\circ}}$, (viii) the time of the closest
approach of the lens trajectory with respect to the source center in
the unit of the crossing time of the source surface
$\mathrm{t_{0}(t_{\star})}$ and (ix) the time of the perihelion
passage in the planetary orbit in the unit of the orbital period
$\mathrm{t_{0,p}(T)}$ respectively. The other parameters are fixed:
the mass of the source star $\mathrm{M_{\star}=M_{\odot}}$, the
radius of the source star $\mathrm{R_{\star}=R_{\odot}}$, the lens
mass $\mathrm{M_{l}=0.3~M_{\odot}}$, the lens and source distances
from the observer $\mathrm{D_{l}=6.5~kpc}$,
$\mathrm{D_{s}=8.0~kpc}$, the limb-darkening coefficients
$\mathrm{c_{1}=0.64}$, $\mathrm{c_{2}=0.032}$, the redistribution
factor $\mathrm{f=2/3}$, the eccentricity of the planetary orbit
$\mathrm{e=0}$, the projection angle of the planetary semi-major
axis with respect to the sky plane $\mathrm{\beta=0^{\circ}}$ and
the scattering albedo in the planet atmosphere
$\mathrm{A_{s}=0.9}$.}\label{tab1}
\end{deluxetable*}

Figure \ref{fig3a} shows a high-magnification microlensing of a CGP
system, while the lens is passing the source surface. The
photometric and polarimetric curves are plotted in the left and
right panels. The simple models without any planet around the source
star are shown with a red solid line, the photometric residual
$\mathrm{\Delta A= (A'-A)/A}$ and the polarimetric residuals in the
polarization components $\mathrm{\Delta P_{1,2}=P'_{1,2}-P_{1,2}}$
(in \%) are plotted in the lower panels. The thin curve in the
right-handed plot shows the intrinsic polarization signal due to the
CGP system which varies with the time. According to this figure, the
maximum photometric deviation due to the stellar occultation is
achieved when the lens approaches the planet while it is crossing
the source center. The maximum polarimetric deviation occurs if the
lens approaches the planet which is passing the edge of the source.
Because the peak of the total stellar flux, considering
limb-darkening, comes from the center of the source, the stellar
polarized flux maximizes at the source edges. The polarimetric
perturbations due to a transiting planet are located over (or
between) the polarimetric maxima. Here, the inclination angle of the
planetary orbit is about $\mathrm{90^{\circ}}$, so that the
reflection and occultation polarizations are separated in time, and
thus just the occultation polarization is magnified. However, when
the inclination angle is lower the reflection and occultation
polarizations can coincide.

\subsection{Lensing of the reflection polarization}
When the lens is passing very close to the source star it may pass
close to the planet and results the reflection polarization signal
to be magnified. In that case, the total Stokes parameters are given
by:
\begin{eqnarray}\label{reflec}
S'_{i}=R^{2}_{\star}S_{i,\star}+ R^{2}_{p}S_{i,r},
\end{eqnarray}
where $\mathrm{i \in (I,Q,U)}$, $\mathrm{S_{i,r}}$ is the total
Stokes parameter due to the reflected light by the planet and is
obtained by integrating over the illuminated part of the planet
surface which is a function of the phase angle. Accordingly, the
magnification factor in a high-magnification microlensing event of a
CGP system, while the lens is magnifying the reflected light by the
planet atmosphere will be:
\begin{eqnarray}
A'=A \frac{1+\delta~S_{I,r}/
S_{I,\star}}{1+\delta~S_{I,r,0}/S_{I,\star,0}}.
\end{eqnarray}
The factors $\mathrm{S_{I,r}/S_{I,\star}}$ and
$\mathrm{S_{I,r,0}/S_{I,\star,0}}$ are proportional to the ratio of
the source flux reflected by the planet atmosphere to the total
source flux, i.e.
$$\mathrm{F^{r}/F^{\star}= A_{g}(\frac{R_{p}}{a})^2\Phi(\alpha),}$$ where $\mathrm{A_{g}}$ is
the geometrical albedo which is the fraction of the incident
reflected light from the planet surface at the phase angle equal to
zero. $\mathrm{\Phi(\alpha)}$ is the phase function and indicates
the fraction of the planet surface that is illuminated by the source
star and can be detected by the observer. In these cases, some
positive or negative deviations can appear in an unperturbed light
curve. Whenever the planet-induced deviation vanishes for a moment
(i.e. $\mathrm{S_{I,r}/S_{I,\star}=S_{I,r,0}/S_{I,\star,0}}$) we
have $\mathrm{A'=A=A_{r}}$, where $\mathrm{A_{r}=S_{I,r}/S_{I,r,0}}$
is the magnification factor of the planet itself by considering its
reflected radiation. Also, the polarization components are given by:
\begin{eqnarray}
P'_{1}&=&P_{1}\frac{1+\delta~S_{Q,r}/S_{Q,\star}}{1+\delta~S_{I,r}/S_{I,\star}},\nonumber\\
P'_{2}&=&P_{2}\frac{1+\delta~S_{U,r}/S_{U,\star}}{1+\delta~S_{I,r}/S_{I,\star}}.
\end{eqnarray}
It is clear that the planet-induced deviations in the polarization
components can also be in the form of positive or negative
deviations with respect to the unperturbed polarization components
of the source star. If at some time between these deviations in each
polarization component the perturbation effect vanishes, then
$\mathrm{P'_{1}=P_{1}=P_{1,p}=S_{Q,r}/S_{I,r}}$ or
$\mathrm{P'_{2}=P_{2}=P_{2,p}=S_{U,r}/S_{I,r}}$. Consequently, when
the lens is passing close to the planet, its reflected light will be
magnified. In the resulting photometric or polarimetric deviations,
the times when these deviations vanish give us some information
about the planet characteristics.

Figure \ref{fig3b} represents a polarimetric microlensing event of a
CGP system for three different values of the planet semi-major axis.
In these events the reflected light by the planet is passing close
to the lens. Here the positive-negative transitions happen in the
second component of the stellar polarization. By increasing the
planet distance from its host star the contribution of the reflected
light significantly decreases.

There is another polarization signal in a micro-lensed CGP system
which is due to being magnified the thermal radiation from the
planet itself. This component is considerably stronger in infrared
bands and depends on the planet's temperature which is given by:
$$\mathrm{T_{p}=T_{\star}(R_{\star}/a)^{1/2}(f(1-A_{b}))^{1/4}},$$
where $\mathrm{f}$ is the redistribution factor which describes the
fraction of re-radiating energy which is absorbed by the planet and
$\mathrm{A_{b}}$ is the bond albedo (or spherical albedo) which is
the total fraction of incident light reflected by a sphere at all
angles. We indicate the geometrical and bond albedos according to
the amount of the scattering albedo $\mathrm{A_{s}}$
\cite{Madhusuhan2012}. In Figure \ref{fig3c} we show a
high-magnification microlensing event of a CGP system in two
different pass bands. In this event, we align the lens trajectory so
that it reaches to the planet when the phase angle $\mathrm{\sim
120^{\circ}}$, i.e. the contributions of the occultation and
reflection polarizations is negligible. According to this figure the
planet-induced perturbation due to the thermal radiation of the
planet increases with increasing the wavelength.

There is a twofold degeneracy in these kinds of events. If
$\mathrm{u_{0}\rightarrow -u_{0}}$ and $\mathrm{i\rightarrow -i}$,
the photometric and polarimetric curves do not change, but the
polarimetric angle
$\mathrm{\theta_{p}'=1/2\tan^{-1}(S'_{U}/S'_{Q})}$ converts to
$\mathrm{180-\theta_{p}'}$. Hence, in these microlensing events
polarimetric observations resolve the twofold degeneracy which is
not doable relying on photometric observations alone. This is one of
the major advantages of polarimetric observation of
high-magnification or caustic-crossing microlensing events.

Let us assume the lens is a binary system also and the source star
is also a CGP system. In these events, while the source star is
passing the caustic its planet can cross the caustic lines several
times and and as a result some periodic perturbations will appear in
the related curves. The orbital period of the CGP system can be
measured through an accurate timing analysis of the residuals. This
method was proposed for measuring the orbital period in the close
binary microlensing events by Nucita et al. (2014). The important
issue in this regard is that the time scale of these planet-induced
perturbations is too short and is of the order of
$\mathrm{t_{p}=t_{E}\rho_{p}}$, where $\rho_{p}$ is the projected
planet radius on the lens plane and normalized to the Einstein
radius. This issue decreases the probability of measuring these
period perturbations. However, this polarimetry time scale will be
considerable in long-duration microlensing events only. Also, in
binary microlensing events of CGP systems, we can obtain some
information about the planet from the times when the planet-induced
perturbations vanishes for a moments in the polarimetry or light
curves, same as single-lens microlensing events.

\begin{figure}
\includegraphics[angle=270,width=8.cm,clip=]{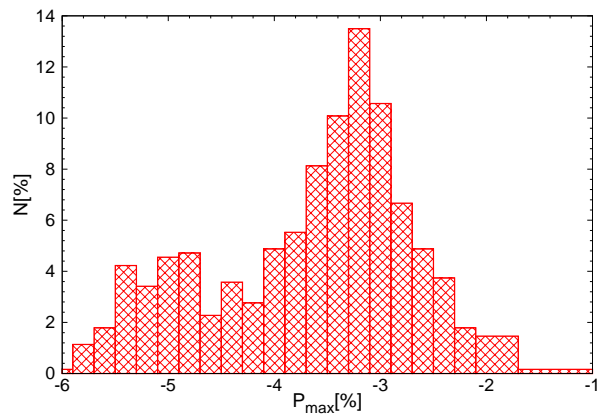}
\caption{The distribution of the maximum polarization signals (in
\%) due to the confirmed CGP systems with semi-major axes with less
than $\mathrm{0.07~A.U.}$.}\label{fig4}
\end{figure}

\section{Observability of CGP systems in high-magnification
microlensing events}\label{five} In this section, we investigate the
polarimetric and photometric efficiencies for detecting the CGP
system signatures in high-magnification microlensing events. We
first study the statistical contribution of the intrinsic
polarization signals due to observed CGP systems which have been
mostly found by the transit method and are available online on the
NASA's exoplanet
archive\footnote{http://exoplanetarchive.ipac.caltech.edu/}. Based
on our selection of CGPs, we simulate high-magnification
microlensing events with source star CGPs and provide detection
probabilities for the photometric and polarimetric signature.

There are about $\mathrm{615}$ confirmed close-in giant planets with
measured semi major-axes $<\mathrm{0.07~A.U.}$ which have been
detected with various methods. We choose these transiting CGP
systems for simulating source stars and randomly pick the
inclination angles $\mathrm{(\beta,i)}$ and calculate their maximum
polarization signals. Figure (\ref{fig4}) shows the distribution of
the maximal detectable polarization signature caused by these CGP
systems. We find that the net polarization cause by these systems is
very low and mostly less than $\mathrm{10^{-3}\%}$. The reflection
polarization signals due to CGP systems can reach
$\mathrm{10^{-3}\%}$ and correspond to the second peak in the Figure
(\ref{fig4}). The occultation polarizations are even weaker and of
the order of $\mathrm{10^{-5}\%}$ which correspond to the first
deviating peak. About $\mathrm{2}$\,\% of CGP systems have a maximum
polarization signal greater than $\mathrm{0.01\%}$.

Then, we simulate high-magnification microlensing events of these
CGP systems to assess the polarimetry and photometry efficiencies
for detecting planet-induced signatures in these events. We envisage
polarimetric observations of these high-magnification microlensing
events are carried out using the FORS2 instrument at the VLT and
according to the observational strategy of this instrument simulate
the observational data points. Further details on simulating
polarimetric microlensing observations with FORS2 can be found in
the work by Sajadian (2016). One example of such a simulated
polarimetric microlensing event of a CGP system is shown in Figure
\ref{fig3d}. In this event, the planet-induced polarimetric
perturbation exceeds the polarimetric accuracy of FORS2 i.e.
$\mathrm{0.1\%}$. The cadence of FORS2 observations is about
$\mathrm{20~min}$ and thus exceeds the time-scale of the
polarimetric perturbation, so that the polarimetric data points can
not cover the second polarimetric peak and resolve the
planet-induced perturbation. The photometric signature of the planet
is very small, but high spatial resolution of the photometric
observations and the short cadence leave the planet-induced
perturbation detectable.

In about $\mathrm{10\%}$ of the simulated events, the lens distance
from the planet decreases to less than $\mathrm{\sim4 \rho_{p}}$.
For $\mathrm{3\%}$ of our simulated events the photometric
signatures of the close-in planet around the source star are
detectable with $\mathrm{\Delta \chi^{2}> 50}$, where
$\mathrm{\Delta\chi^{2}=\chi^{2}_{p}-\chi^{2}_{s}}$ is the change in
$\mathrm{\chi^{2}}$ for a fitted microlensing light curve with and
without CGP system orbiting the source. The polarimetric
perturbations were not detectable in any simulated event for that
criterion. It means that the polarimetric efficiency for detecting
CGP systems in high-magnification microlensing events is less than
$0.1\%$ for one state-of-the-art instrument with a corresponding
polarimetric accuracy of $\mathrm{0.1\%}$ and an observational
cadence about $\mathrm{20~min}$. Consequently, as expected the
photometry observations are more efficient for detecting CGP systems
around the source stars than the polarimetry observations.

The small polarimetry efficiency for detecting CGP systems orbiting
the source stars is not only owing to the low polarimetry precision
of FORS2 but also more due to the long necessary observational time
for taking one polarimetry data point by FORS2. Regarding the second
issue, the FORS2 cadence (about $\mathrm{20~min}$) is of the other
of the time scale of the planet-induced polarimetry perturbations
(i.e. $t_{p}=\rho_{p}t_{E}$) which causes a lack of enough number of
the polarimetry data points to cover these perturbations, as shown
in Figure \ref{fig3d}. It means that if several high resolution
polarimeters observe high-magnification microlensing events
synchronously, the polarimetry efficiency for detecting this kind of
perturbations significantly increases.

However, the advantages of potential high-resolution and
short-cadence polarimetry observations in near future from
high-magnification microlensing events of CGP systems will be (i)
resolving the twofold degeneracy which is not resolvable by doing
photometry observations alone and (ii) obtaining some extra
information about the planet through the times in which the
planet-induced perturbations vanishes for an instance, as explained
in the previous section.

\section{Conclusions}\label{five}
More than $\mathrm{600}$ close-in giant planetary (CGP) systems with
semi-major axes less than $\mathrm{0.07~A.U.}$ have been confirmed
up to now. Most of the detected systems are closer than
$\mathrm{1~kpc}$ from us and were often detected by the transit
method. CGP systems located in the Galactic bulge distance can be
detected only through gravitational microlensing. Finding such
systems in the Galactic bulge would help us to study the
distribution of such systems in different environments and
understand the environmental effect on the formation of these
systems.

One method for detecting CGP systems through gravitational
microlensing is polarimetry observations. In that case, planetary
signatures can be magnified beyond a common detection threshold if
(i) a high-magnification microlensing event occurs or (ii) if the
system crosses a caustic line. Unmagnified CGP systems will also
lead to a net polarization signal, albeit too small to be measured
directly. During a high-magnification or caustic-crossing
microlensing event, the undetectable polarization signal can be
magnified which causes deviations in their polarimetric curves and
was discussed in this work. The polarization signal itself is caused
by the stellar occultation by the planet or by the reflection of the
host star's light from the planet surface and strongly depends on
the orbital radius, the density and surface area of the planet and
the chemical components in the planetary atmosphere. Polarization by
reflection exceeds the signal caused by the primary transit by about
two orders of magnitude.

While the lens is crossing the source star surface, the occultation
polarization is likely to be magnified and when the lens is
sufficiently close to the source star and out of its surface, the
reflection polarization can be magnified, too. The resulting
perturbations do not have commonly identifiable shapes, because lens
and planet trajectories can come in a variety of configurations. But
in all cases of these photometric or polarimetric perturbations.
finding the times when the polarimetric or photometric perturbation
vanishes gives some information about the planet.

We have also investigated the observability of detecting photometric
and polarimetric deviations in the related curves of
high-magnification microlensing events of a sample of confirmed CGP
systems in NASA's exoplanet database. We conclude that the
photometric efficiency for detecting these signatures is about
$\mathrm{3\%}$, whereas the polarimetric efficiency is less than
$\mathrm{0.1\%}$ which is comparable to state-of-the-art instruments
with a polarimetric accuracy $\mathrm{0.1\%}$. The small polarimetry
efficiency for detecting CGP systems orbiting the source stars is
owing to the low polarimetry precision of FORS2 and rather due to
the long necessary observational time for taking one polarimetry
data point by FORS2. If several high resolution polarimeters observe
high-magnification microlensing events synchronously, the
polarimetry efficiency for detecting this kind of perturbations
significantly increases. However, the advantages of high-resolution
and short-cadence polarimetry observations in near future from
high-magnification microlensing events of CGP systems will be (i)
resolving the twofold degeneracy which is not resolvable by doing
photometry observations alone and (ii) getting some extra
information about the planet through the times in which the
planet-induced perturbations vanishes for an instance.

\begin{acknowledgments}
We are thankful from N. Kostogryz for providing her results of
simulating the center-to-limb polarization profile in continuum
spectra of low-mass stars. We would like to thank the anonymous
referee for his/her helpful comments and suggestions.
\end{acknowledgments}

\end{document}